              \documentstyle[prl,preprint,aps,epsf]{revtex}  
\newcommand{\postscript}[2]
 {\setlength{\epsfxsize}{#2\hsize}
  \centerline{\epsfbox{#1}}}
\def\dm2{ \Delta m^2 }
\def\s22t{ \sin^22\theta }

\def\be{ \begin{equation} }
\def\ee{ \end{equation} }
\def\ba{\begin{array} }
\def\ea{\end{array} }
\def\bea*{ \begin{eqnarray*} }
\def\eea*{ \end{eqnarray*} }

\begin{document}

\renewcommand{\baselinestretch}{1.0}

\title{
\hfill               {\normalsize \today
\\       \hfill                        UPR--0632T}
\\
         Implications of High Precision Experiments
         and the CDF  \\ Top Quark Candidates
}
\author{
   Jens Erler and Paul Langacker
}

\address{
   Department of Physics,
   University of Pennsylvania, Philadelphia, PA 19104, U.S.A.
}

\maketitle

\begin{abstract}
We discuss the consequences of recent experimental results from CDF, SLC,
LEP and elsewhere for the Standard Model and for new physics.
A global fit to all indirect precision data yields
$m_t = 175 \pm 11^{+17}_{-19}$~GeV,
$\sin^2 \hat\theta_{\overline {MS}} = 0.2317 (3) (2)$, and
$\alpha_s = 0.127 (5) (2)$, where the central values are for $M_H = 300$ GeV
and the second uncertainties are for $M_H \rightarrow 1000$ GeV (+) and
60 GeV ($-$). The $m_t$ value is in remarkable agreement with the value
$m_t=174 \pm 16$ GeV suggested by the CDF candidate events. There is a slight
preference for a light Higgs with $M_H < 730$ (880) GeV at 95\% c.l.
if the CDF $m_t$ value is (not) included. The sensitivity is, however,
due almost entirely to the anomalously large observed values for the
$Z \rightarrow b\bar{b}$ width and left-right asymmetry.
The value of $\alpha_s$ (from the lineshape) is clean theoretically
assuming the Standard Model, but is sensitive to the presence of new physics
contributions to the $Z \rightarrow b\bar{b}$ vertex.
Allowing a vertex correction $\delta_{b\bar{b}}^{\rm new}$ one obtains
the significantly lower value $\alpha_s = 0.111 \pm 0.009$, in better
agreement with low energy determinations, and  $\delta_{b\bar{b}}^{\rm new}
= 0.023 \pm 0.011$.
There is now enough data to perform more general fits to parameters
describing new physics effects and to separate these from $m_t$ and
$M_H$. Allowing the parameter $\rho_0$, which describes sources of $SU(2)$
breaking beyond the Standard Model, to be free one finds
$\rho_0=1.0012 \pm 0.0017 \pm 0.0017$,
remarkably close to unity.
One can also separate the new physics contributions to the oblique parameters
$S_{\rm new}$, $T_{\rm new}$ and $U_{\rm new}$, which all take values
consistent with zero. The effects of supersymmetry on the determination of
the Standard Model parameters are discussed.

\end{abstract}



\renewcommand{\baselinestretch}{2.0}

\section{INTRODUCTION:}

In April 1994 the CDF collaboration~\cite{CDF} reported
evidence for top quark production at the Tevatron. The collected events
are consistent with a top quark (pole) mass
$m_t = 174 \pm 16$~GeV.
Electroweak radiative corrections to Standard Model (SM) observables
have large $m_t$ effects, so that its direct determination is of upmost
importance. Significant indirect bounds on the Higgs mass $M_H$ can
only be obtained after $m_t$ is known independently. Top quark
effects impede the setting of bounds on or the discovery of new physics
beyond the Standard Model from precision observables. In this paper we study
the implications of the $m_t$ range suggested by the CDF candidate events for
high precision measurements. For comparison, we carry out analyses of the
Standard Model parameters with and without the CDF constraint.

Only one month prior to the
announcement from Tevatron, the SLD collaboration~\cite{SLD2} at SLAC
published a precise measurement of the left-right asymmetry $A_{LR}$.
The polarization of the SLC electron beam was increased from
22\% to 63\% and the number of $Z$ events increased by a factor of 5
compared to the 1992 run. At about the same time the LEP groups~\cite{LEP}
presented a first analysis of their 1993 data. The integrated
luminosity in 1993 amounted to 40 ${\rm pb}^{-1}$, a number only
slightly smaller than the integrated luminosities from the previous years
combined. Thus, the new (and preliminary) data contribute with a
high statistical weight. Moreover, systematic uncertainties were significantly
reduced, most notably in the total $Z$ width, increasing the significance of
the 1993 run even further.

The experimental results are summarized in table~\ref{tblexp}, together with
the SM expectations using the global best fit values $m_t = 175 \pm 11$ GeV
(for $M_H = 300$ GeV) and $\alpha_s = 0.127 \pm 0.005$ (see below).
The three errors in the SM predictions correspond respectively to
(1) the uncertainties in $M_Z$ and $\alpha (M_Z)$,
(2) the (correlated) uncertainties from $m_t$ and $M_H$ (which can vary from
60 to 1000 GeV, with a central value of 300 GeV), and (3) the uncertainty
in $\alpha_s$. $\sigma_{\rm had}^0$ is the bare hadronic peak
cross section, i.e., the cross section at $\sqrt{s} = M_Z$ after correcting
for photonic contributions. Similarly, $A_{FB}^{0f} = {3\over 4}
A_e^0 A_f^0$ is the bare
forward-backward asymmetry for $e^+ e^- \rightarrow Z \rightarrow f\bar{f}$,
$A_{FB}^{0l}$ is the asymmetry for charged leptons assuming family
universality (after correcting for $m_\tau$), $\bar{s}^2_e (Q_{FB})$ is the
effective weak angle determined from the hadronic charge
asymmetry, and $A_{LR}^0 = A_e^0$
is the bare left-right polarization asymmetry. The quantity
$A^0_f$ for flavor $f$ is defined by\footnote{$\bar{s}_f^2$
and $\bar{g}_{V,A_f}$ are flavor dependent effective
mixing angles and vector (axial-vector) couplings. They include propagator and
vertex
corrections evaluated at $s=M_Z^2$.}
\be
   A_f^0 =
   \frac {2 \bar{g}_{V_f} \bar{g}_{A_f}}{\bar{g}_{V_f}^2 + \bar{g}_{A_f}^2}.
\ee
$N_\nu$ is the number of active neutrinos (with $m_\nu \leq M_Z/2$);
$Q_W$ is the effective weak charge in atomic
parity violation~\cite{atompar}; $g_{V,A}^{\nu e}$ are effective 4-Fermi
couplings for neutrino electron scattering~\cite{nueel}; and $s^2_W =
1-M_W^2/M_Z^2$
is the on-shell weak mixing angle from deep inelastic neutrino nucleon
scattering~\cite{nuenuc}. Our fits
include the older low energy data~\cite{olddata} as well.

The agreement between theory and
experiment is generally excellent, with the exceptions\footnote{Also,
the individual forward-backward $\tau$ asymmetry, $A_{FB}^{0\tau} =
0.0228 \pm 0.0028$ is 2.8 $\sigma$ above SM expectations
and renders the test of lepton
universality only moderately successful.
If the FB asymmetries into the 3 lepton species are
used, one finds that universality is excluded at the
93\% c.l.~\cite{Teupitz}.
There is also a direct discrepancy between $A_{FB}^{0\tau}$ and
$A_\tau^0({\cal P}_{\tau})$ (which is consistent with the SM)
of about 2.5 $\sigma$.} of $A_{LR}^0$ and $R_b$. Of course, given
the plethora of precision measurements one expects deviations at some
level, and it would be premature to take the deviating quantities described
below as serious problems for the SM. However, these quantities,
especially $R_b$, have a significant effect on some of the
conclusions, so it is worthwhile to comment on them.

The left-right asymmetry, $A_{LR}$, as measured at SLC, is the most precise
single determination of the effective weak angle
$\bar{s}_e^2$~\cite{SLD2}. SLD quotes $A_{LR} (\sqrt{s}
= 91.26\; {\rm GeV}) = 0.1628 \pm 0.0076$. Correcting for photon
exchange, electroweak interference, and initial-state radiation yields
\be
   A_{LR}^0 = A_e^0  = 0.1656 \pm 0.0076.
\ee
This corresponds to
\be
\label{sineff}
   \bar{s}_e^2 = {1\over 4 |Q_e|}
   (1 - {\bar{g}_{V_e}\over \bar{g}_{A_e}}) = 0.2292 \pm 0.0010.
\ee
Inclusion of the 1992 result~\cite{SLD1},
$A_{LR}^0 = 0.100 \pm 0.044$, yields combined values $A_{LR}^0 =
0.1637 \pm 0.0075$ and $\bar{s}_e^2 = 0.2294 \pm 0.0010$, which
is about 2.5 $\sigma$ deviation from global SM fits
and at least 2 $\sigma$ from the values
derived from LEP asymmetries.
The 1992 + 1993 result (combined with the LEP value of $M_Z$)
yields $m_t = 251^{+24}_{-26}$ GeV for $M_H = 300$ GeV.
Relaxing the universality assumption or even allowing for the
most general fermion couplings to the $Z$ do not considerably improve the
goodness of the fits~\cite{Teupitz}.
Also, using results from reference~\cite{LLM} it was
argued in~\cite{Teupitz} that no kind of new physics can account for
the SLD result without simultaneously conflicting one or several other
observables, most notably the $W$ mass. Thus assuming that these experiments
are completely governed by $\gamma$ and $Z$ amplitudes we look at a direct
experimental conflict. One possible loophole, namely the presence of new
effective interactions which contribute significantly to the lineshape
and asymmetries, is discussed in reference~\cite{4Fermi}.

The $Z \rightarrow b\bar{b}$ vertex has long been
advertized~\cite{Jegerlehner} as the ``ideal top mass meter''
since it is virtually
independent of $M_H$, and the terms quadratic in $m_t$ enter in a different
way than in $\hat\rho \equiv M_W^2/M_Z^2 \cos^2\hat{\theta}_{\overline {MS}}$
which governs other electroweak observables.
At the same time it is sensitive to many kinds of physics beyond the SM.
The LEP groups obtain~\cite{LEP}
\be
   R_b = \frac{\Gamma (b \bar{b})}{\Gamma ({\rm had})} = 0.2202 \pm 0.0020,
\ee
from a fit with $R_c = \Gamma (c\bar{c})/\Gamma ({\rm had})$ left
free\footnote{
We will always use the experimental values of $R_b$ and $R_c$ with
their correlation of $-0.4$~\cite{LEP}. Alternatively, one could use
the value $R_b = 0.2192 \pm 0.0018$, obtained~\cite{LEP} by fixing $R_c$ to its
SM value of 0.171. We have checked that the two methods yield
virtually identical results.}.
This is 2.3 standard deviations from the SM prediction
$R_b = 0.2156 \pm 0.0004$. $R_b$ drives the fits to smaller values of
$m_t$, independent of $M_H$. Due to the correlation of top and Higgs effects
in the $\hat\rho$ parameter, this in turn favors smaller values of $M_H$.

With the possible exception of these measurements, experiments and the
minimal Standard Model are in spectacular agreement with each other.

\section{RADIATIVE CORRECTIONS:}

One of the goals of the $Z$ factories at SLAC and CERN is to test
electroweak theory at the quantum level. $m_t$, $M_H$ and $\alpha_s$
enter only through radiative corrections. They are obscured
by pure QED corrections, which are large but calculable and under control,
with the possible exception of small-angle ${\cal O} (\alpha^2)$
Bhabba scattering. Experimenters usually present data with all QED corrections
other than final state radiation removed.

The hadronic contribution to the vacuum polarization~\cite{Jegerlehner}
induces an uncertainty
of 0.0003 in $\sin^2 \theta_W$. Omitting this error in the global fits
can change the extracted value of $m_t$ by about 3 GeV.

QCD corrections are calculated~\cite{QCD3} and included up to
${\cal O} (\alpha_s^3)$. We did not include ${\cal O} (\alpha_s^4)$
corrections, which are estimated to contribute with a negative sign
and with about 0.4 MeV to the hadronic $Z$ width~\cite{QCD4}, corresponding
to an additional uncertainty of 0.001 in $\alpha_s$.
As pointed out in~\cite{QCDT2} the ${\cal O} (\alpha_s^2)$ corrections to the
vector and axial vector parts of the partial $Z$ width into $b$ quarks
exhibit different dependences on $m_t$. They are important and
included along with the analogous results for the ${\cal O} (\alpha_s^3)$
corrections~\cite{QCDT3}. Higher order QCD corrections proportional to
$m_b^2/M_Z^2$ are incorporated as well~\cite{QCDB}.

As for the electroweak sector, full one loop corrections are taken into
account. Due to the heavy top quark, two loop effects of
${\cal O} (\alpha^2 m_t^4)$
are included with their full $M_H$ dependence~\cite{BBCCV},
as well as ${\cal O} (\alpha \alpha_s m_t^2)$ corrections to the $\rho$
parameter~\cite{QCDEWR} and to the $Z \rightarrow b\bar{b}$
vertex~\cite{QCDEWB}. Threshold effects corresponding to
${\cal O} (\alpha \alpha_s^2 m_t^2)$ corrections are incorporated by making
use of the detailed work of Fanchiotti, Kniehl and Sirlin~\cite{FKS}.
They can also be estimated~\cite{SV} by employing $\alpha_s (0.15 m_t)$
rather than $\alpha_s (m_t)$. The numerical difference between the two
approaches is negligible, and either way threshold effects increase
the extracted top mass by about 3 GeV.

In practice we used the routine ZFITTER~\cite{ZFITTER} for the calculation
of form factors. The improved Born formulae were then dressed with the
aforementioned QED and QCD corrections for the $Z$ partial widths.
The agreement with ZFITTER version 4.6 is excellent with differences being at
the 0.1 MeV level in the total $Z$ width. For the most important form
factor, $\hat{k}_e (M_Z^2)$ (see eq.~(\ref{sinsin}) below), we used
the update by Gambino and Sirlin~\cite{GS}.

On the quantum level the exact definition of the weak mixing angle becomes
ambiguous. Besides the conceptually most simple on-shell definition,
$s^2_W \equiv 1 - M_W^2/M_Z^2 = 0.2243 \pm 0.0012$,
there are two other definitions\footnote{Yet another definition,
$s_{M_Z}^2 = 0.2312 \pm 0.0003$, is obtained by removing the $m_t$
dependence from the expression for $M_Z$~\cite{NOV}. The $m_t$
uncertainty reenters when other observables are expressed in terms of
$s_{M_Z}^2$. The various definitions are further discussed
in~\cite{PDG26}.} which are numerically very close to each other.

One is based on the coupling constants,
$\tan \hat{\theta}_W (M_Z) \equiv g^\prime/g$, which are radiatively
corrected according to the ${\overline {\rm MS}}$ prescription. This
makes the ${\overline {\rm MS}}$ quantity $\sin^2 \hat{\theta}_W \equiv
\hat{s}^2_Z = 0.2317 \pm 0.0004$ particularly
convenient for GUT predictions and
insensitive to new physics. However, it is a quantity designed by theorists
and is not related simply to any single observable. Rather, it is best
determined by a global fit. Also, there are variant forms of
$\hat{s}^2_Z$ which differ in the treatment of heavy top
quark effects. A variant~\cite{DFS} in which the heavy top quark
is not decoupled is a few times $10^{-4}$ larger~\cite{GS} than the one
introduced in~\cite{MR}. In the latter, which is used
here, the ln $m_t$ effects in $\gamma$ -- $Z$ mixing are decoupled,
so that the $Z$-pole asymmetries are essentially independent of $m_t$.

The other is the effective mixing angle defined in
eq.~(\ref{sineff}), with analogous definitions for other flavors.
It is defined through observables (the $Z$-pole asymmetries), which makes it
conceptually simple, but for the exact relation to other quantities
a computer code is needed
due to the need to compute three point functions. That also makes it
difficult to relate $\bar{s}^2_e$ to non $Z$-pole observables.

The two definitions above share a smaller sensitivity to $m_t$ compared
to the on-shell $s^2_W$. For the relation between $s^2_W$ (or $M_W$),
$\hat{s}^2_Z$, and $M_Z$ we rely on
reference~\cite{FKS}\footnote{We are indebted to Bernd Kniehl, who made
his computer code on which the numerical results of ref.~\cite{FKS} are based
available to us.}. $\hat{s}^2_Z$ and $\bar{s}^2_e$ are related by
\be
\label{sinsin}
   \bar{s}^2_e = \hat{s}^2_Z {\rm Re}\; \hat{k}_e (M_Z^2),
\ee
with the form factor ${\rm Re}\; \hat{k}_e (M_Z^2)$ from reference~\cite{GS}.
Relation~(\ref{sinsin}) is a very good approximation due to the smallness
of ${\rm Im}\; \hat{k}_e (M_Z^2)$. For $m_t$ in the relevant range,
(\ref{sinsin}) implies
\be
   \bar{s}_e^2 \sim \hat{s}^2_Z + 0.00028.
\ee

\section{FIT RESULTS:}

We regard the deviations in some of the observables as consistent with
statistical fluctuations and have therefore refrained from using scale factors
to increase error
bars, and instead simply combined the data\footnote{A scale factor of 2.2
for the uncertainties in $A_e$ from $A_{LR}$ and ${\cal P}_\tau$,
as suggested by the Particle Data Group~\cite{PDG}, would decrease
the value of $m_t$ predicted by the indirect data by 5 GeV.}.

Table~\ref{tblres} summarizes the results of various fits to
$\hat{s}^2_Z$, $\alpha_s (M_Z)$, and $m_t$
based on different data sets. The central values\footnote{The predictions
in Table~\ref{tblexp}, especially
for $\bar{s}^2_e (Q_{FB})$, differ slightly from the values at
the best fit point, because the former use the central value
of $\alpha^{-1} (M_Z) = 127.9 \pm 0.1$~\cite{FKS},
incorporating the $\pm 0.1$ in the first listed uncertainty, while
the best fit occurs at $\alpha^{-1}(M_Z) = 128.0$.} correspond to
$M_H = 300$ GeV and the second errors indicate the results for
$M_H =1000$ GeV (+) and $M_H = 60$ GeV $(-)$. The increase in $\chi^2$
when changing $M_H$ from 60 to 1000 GeV,
$\Delta \chi^2_H \equiv \chi^2 (1000) - \chi^2 (60)$, is also indicated.
The first row is the fit to all indirect precision data\footnote{The
overall $\chi^2$ is 181 for 206 d.o.f. This is rather low (mainly due
to the earlier neutral current data), but statistically acceptable:
the probability of $\chi^2 \leq 181$ is 10\%. The correlation
coefficients are $\rho_{\hat{s}^2_Z \alpha_s} = 0.30$,
$\rho_{\hat{s}^2_Z m_t} = -0.67$, $\rho_{\alpha_s m_t} = -0.20$. The
correlations for the other data sets are similar.}. The prediction $m_t
= 175 \pm 11^{+17}_{-19}$ GeV is in remarkable agreement with the CDF
value $174 \pm 16$ GeV. Not surprisingly, including the CDF value as
an additional constraint (second row) has little impact on the global
fit within the Standard Model. It will, however, be of great
importance in the non-Standard Model fits. The third row is a fit in
which the indirect data is combined with the additional constraint
$\alpha_s (M_Z) = 0.116 \pm 0.005$ obtained from data other than the
$Z$ lineshape~\cite{PDG}. As expected, the extracted $\alpha_s$ (which
can be regarded as a simultaneous fit to the lineshape and other
$\alpha_s$ data) is somewhat lower than the value from the lineshape
alone. The other rows are fits to subsets of the data, which show
the sensitivity to the various inputs. From the fourth row (LEP + low
energy) we see that the predicted $m_t$ is 7 GeV lower without
$A_{LR}$ from SLD, while when averaging $A_e^0$ from $A_{LR}^0$ and from
${\cal P}_\tau$ with a scale factor of 2.2 (fifth row) it is lower by 5 GeV.
The results from the $Z$-pole (LEP + SLD), LEP, and SLD + $M_Z$ are also
shown. The large value of $m_t$ in the last case reflects the high
value of $A_{LR}^0$.

It is useful to compare these results with the fits performed by the
LEP electroweak working group~\cite{LEP}. Their fits for $Z$-pole, $M_W$, and
recent neutrino data (which corresponds roughly to our first ``All
indirect'' fit), as well as to the LEP data are displayed in the last
two rows of Table~\ref{tblres}. The agreement between their results
and ours is excellent, with the small (correlated) differences in
$\alpha_s$ and $m_t$ a reflection of the completely independent
implementation of radiative corrections.

There is a slight preference for a light Higgs (as
is predicted in the minimal supersymmetric extension of the SM) but it is weak
statistically. Combining all
indirect data we can set an upper limit on $M_H < 570$ (880) GeV at the
90\% (95\%) c.l. Adding the CDF result this limit is strengthened to
$M_H < 510$ (730) GeV.
Moreover, the preference is driven mainly by the anomalous values of $R_b$
and $A_{LR}$. Removing them from the data set leads to an almost flat
$\chi^2$ distribution with respect to $M_H$, as is shown in Figure~\ref{Higgs}.
Hence, caution is called for in drawing any conclusion on $M_H$ from the
present data.

In the context of the
SM, $R = \Gamma ({\rm had}) / \Gamma (l^+ l^-)$ is a theoretically
clean measurement of $\alpha_s (M_Z)$. In the presence of new physics which
increases the hadronic or $b\bar{b}$
event sample, however, $R$ loses its sensitivity
to the strong coupling constant. Similar remarks hold for
$\Gamma_Z$, which is also sensitive to $\alpha_s$. In 1993 the LEP groups
collected data at the $Z$ peak and at $\pm 1.8$ GeV away from the peak. That
allowed for a precise measurement of the $Z$ lineshape. The extracted
$\Gamma_Z$ is about 1 standard deviation higher than in 1992 and the error
decreased by almost 50\%. It should be noted, however, that a lineshape scan
involving only 3 scan points cannot by itself
be sensitive to any non-$Z$ pole contribution to the cross section.
An overconstrained lineshape fit is only possible when the lower statistics
1990/91 scan is included.

The extracted value of $\alpha_s = 0.127 \pm 0.005 \pm 0.002$ is consistent
with the LEP jet event shape analysis, which yields
$\alpha_s = 0.123 \pm 0.006$ and with the value $0.122 \pm 0.005$ from the
hadronic $\tau$ decay fraction~\cite{PDG,Bethke,BC}.
It is also in perfect agreement with
grand desert SUSY-GUT expectations, favoring
$\alpha_s = 0.127 \pm 0.002 \pm 0.008$, where the first error is due to $m_t$
and $M_H$ and the second arises from the lack of knowledge of the sparticle
and GUT particle spectra (thresholds) and from the unknown effects of possible
nonrenormalizable operators~\cite{Nir}. It is, however, significantly higher
than the values~\cite{PDG,Bethke,BC} obtained from deep inelastic neutrino
and lepton scattering ($0.112 \pm 0.005$), from $J/\Psi$ and $\Upsilon$ decays
($0.113 \pm 0.006$), and from
determinations relying on lattice calculations of the
charmonium ($0.110 \pm 0.006$)~\cite{ElKhadra} and bottomonium
($0.115 \pm 0.002$)~\cite{Davies}
spectra. (The lower energy determinations must of course be
extrapolated to $M_Z$.) As will be discussed in the next section, if one
allows for the possibility of new physics in the $Zb\bar{b}$ vertex to account
for $R_b$, the extracted value of $\alpha_s$ decreases to a lower value
($0.111 \pm 0.009 \pm 0.001$) consistent with these latter values.

\section{New Physics:}

In the context of the Minimal Supersymmetric Standard Model (MSSM) and all its
extensions under discussion, the lightest Higgs eigenstate is known to be
light, in the range 60 GeV $ < M_H < $ 150 GeV. Taking as a central value
$M_H = M_Z$ the extracted top mass is lowered to
$m_t = 160^{+11 \; +6}_{-12 \; -5}$
GeV because of the strong $m_t$ -- $M_H$ correlation\footnote{The other
parameters are $\hat{s}_Z^2 = 0.2316(3)(1)$ and $\alpha_s = 0.126(5)(1)$.}.
In most parts of the MSSM parameter space, i.e., whenever the sparticles and
second Higgs doublet are much heavier than $M_Z$, the decoupling theorem
applies and the only signs of supersymmetry in the precision observables
are a light Higgs and the {\em absence\/} of deviations from the SM.

On the contrary, in Extended Technicolor (ETC) and compositeness models we
expect a variety of effects, most notably the observation of large
Flavor Changing Neutral Currents (FCNC). As an example, in models with
composite fermions,
the effective 4-Fermi operators formed by constituent
interchange have to be strongly suppressed. If we call the compositeness
scale $\Lambda$, so that a 4-Fermi operator takes the form
\be
   L=\pm \frac{4 \pi}{\Lambda^2} \bar{f_1}\Gamma f_2 \bar{f_3}\hat{\Gamma}f_4,
\ee
from FCNC we must require $\Lambda \geq {\cal O}(100\; {\rm TeV})$
unless a fine tuning is invoked. Even then,
atomic parity violation experiments set lower limits~\cite{leptoquarks}
$\Lambda \geq {\cal O}(10\; {\rm TeV})$. They are expected to be
increased to ${\cal O}(40\; {\rm TeV})$ with upcoming experiments.
In ETC models, again contrary to observations, $R_b$ is usually expected to be
below the SM value. Other predictions, at least of the simplest versions
based on scaled-up QCD dynamics, are
$S_{\rm new} > 0$ and $T_{\rm new} \neq 0$, neither of which
are in conformity with the data (see below).

It has become customary to use three quantities, e.g.\ $S$, $T$, and
$U$~\cite{STU},
to parametrize the flavor independent oblique radiative corrections,
or a single parameter, $\rho_0$, to characterize new sources of $SU(2)$
breaking. In
addition, new physics may in particular affect the $Z \rightarrow b\bar{b}$
vertex. In the past it was difficult to disentangle
possible new physics effects from the dominant top mass contributions.
With the new CDF result for $m_t$, however, it is now possible to clearly
determine the new physics effect from the data.

We introduce new variables $S_{\rm new}$, $T_{\rm new}$ and $U_{\rm new}$ via
\be
\ba{l}
   S = S_{\rm new} + S_{m_t} + S_{M_H}, \\
   T = T_{\rm new} + T_{m_t} + T_{M_H}, \\
   U = U_{\rm new} + U_{m_t},
\ea
\ee
where $S_{m_t}$ and $S_{M_H}$ are respectively the $m_t$ and $M_H$
contributions to $S$, and similarly for $T$ and $U$. The effects of $S$,
$T$ and $U$ on the SM expressions for observables are given in~\cite{PDG27}.

We parametrize new physics entering the $Z \rightarrow b\bar{b}$
width by $\delta_{b\bar{b}}^{\rm new}$, defined by~\cite{ABC}
\be
\label{zbb}
  \Gamma_{b\bar{b}}=\Gamma_{b\bar{b}}^{\rm SM}(1+\delta_{b\bar{b}}^{\rm new}).
\ee
$\delta_{b\bar{b}}^{\rm SUSY}$ for the MSSM was computed
in references~\cite{Zbbsusy} and found to be positive or negative depending on
the part of parameter space considered. If the experimental deviation of
$\Gamma_{b\bar{b}}$ from the SM is to be explained by supersymmetry, then there
must be one sparticle light enough to be detected soon.
In typical ETC models and in particular in the explicit model by Appelquist
and Terning~\cite{AT} in which the ETC gauge bosons are weak singlets
and no fine tuning occurs, $\delta_{b\bar{b}}^{\rm ETC}$ is negative and
proportional to $m_t$~\cite{ZbbETC1}.
In models in which the ETC gauge bosons are weak doublets,
the sign in the corresponding contribution to $\delta_{b\bar{b}}^{\rm ETC}$
is reversed. However, there is a competing effect from weak gauge boson mixing
which tends to cancel the former. Hence, a model independent statement about
the sign and the size of $\delta_{b\bar{b}}^{\rm ETC}$ in this class of models
is not possible~\cite{ZbbETC2}. $\delta_{b\bar{b}}^{\rm new}$ may also be used
to set limits on the admixture of extra particles such as an additional
(SU(2) singlet) $D_L$ quark,
since $b_L - D_L$ mixing reduces $\Gamma_{b\bar{b}}$~\cite{LLM}.
In the fits $\delta_{b\bar{b}}^{\rm new}$
affects and is determined by $R_b$, $R$,
$\Gamma_Z$ and $\sigma^0_{\rm had}$.

One can also study the quantity
\be
   \rho_0 \equiv \rho_0^{\rm tree} + \rho_0^{\rm loop}
          \equiv \rho_0^{\rm tree} + \alpha T_{\rm new},
\ee
which describes non-standard sources of (vector) SU(2) breaking.
If $\rho_0 \neq 1$ one has to replace
\be
\ba{rcl}
   M_Z &\rightarrow& {1 \over \sqrt{\rho_0}} M_Z^{\rm SM}, \\
   \Gamma_Z &\rightarrow& \rho_0 \Gamma_Z^{\rm SM}, \\
   {\cal L}_{NC} &\rightarrow& \rho_0 {\cal L}_{NC}^{\rm SM},
\ea
\ee
where ${\cal L}_{NC}$ is a neutral current amplitude (effective Lagrangian).
$\rho_0^{\rm tree}$ differs from unity in the presence of Higgs triplets
or higher Higgs representations,
\be
   \rho_0^{\rm tree} = \frac{\sum_i (t_i^2 - {t_3}_i^2 + t_i)
   |\langle\phi_i\rangle|^2}{\sum_i 2 {t_3}_i^2 |\langle\phi_i\rangle|^2},
\ee
where $t_i$ and ${t_3}_i$ are the weak isospin and its third component
of the neutral Higgs field $\phi_i$.
$\rho_0^{\rm loop}$ gets a positive definite\footnote{Non-degenerate multiplets
involving Majorana fermions or scalars with non-zero vacuum expectation values
can give contributions of either sign.} contribution in the presence
of additional non-degenerate scalar or fermion doublets,
\be
   \rho_0^{\rm loop} = 1 + {3 G_F \over 8 \sqrt{2} \pi^2}
                           \sum_i {C_i \over 3} F(m_{1i},m_{2i}),
\ee
where $C_i$ is the color factor and $F$ a function of the internal particle
masses.
In typical (level~1) superstring models and in grand desert SUSY-GUT models
$\rho_0$ is close to 1, while $\rho_0 \neq 1$ in most compositeness models.
Allowing $\rho_0 \neq 1$ is a special case of the $S_{\rm new}$, $T_{\rm new}$,
$U_{\rm new}$ parametrization, corresponding to $S_{\rm new} = U_{\rm new} =0$,
and $\rho_0 = 1 + \alpha T_{\rm new}$. (Higher dimensional Higgs
representations are technically not included in the standard definition of
$T_{\rm new}$. In practice, however, they cannot be distinguished from oblique
contributions from the precision observables alone, so we will include
both in our definition of $T_{\rm new}$.)

With the CDF result we can now simultaneously determine $\hat{s}^2_Z$,
$\alpha_s$ and $m_t$ as well as a variety of parameters describing
physics beyond the SM. Table~\ref{NPfits} shows the results of various fits
allowing for different parameters left free. In these fits $m_t$ comes mainly
from the direct CDF result and $\hat{s}^2_Z$ from the asymmetries, and
since (given the value of $M_Z$) they are consistent with each other in the SM,
they are largely insensitive to the presence of the new physics parameters.
With them $T_{\rm new}$ can be extracted from $\Gamma_Z$, $S_{\rm new}$
is determined by $M_Z$ and $U_{\rm new}$ from $M_W$.
When $\delta_{b\bar{b}}^{\rm new}$
is left free the large observed value of $R_b$ drives it significantly to
positive values and contributes this way to the hadronic partial $Z$ width.
Although this increase in $\Gamma ({\rm had})$ is only at the per mille level
this has a large effect on the extracted value\footnote{The same effect
on $\alpha_s$ would be obtained if one used the measured value of $R_b$
rather than the Standard Model formula in the expression for
$\Gamma ({\rm had})$.} of $\alpha_s$ which is mainly
determined by $R$: it is driven to lower values, though with a larger error.
The reason is that this determination is only
a loop effect. The other observables, however, are only weakly correlated.

Note, that before the announcement of the CDF top quark candidates the oblique
parameters $S$, $T$ and $U$ could only be discussed relative to some arbitrary
reference value of $m_t$. In particular, it was difficult to separate the
effects of a heavy top on $T$ (or equivalently on $\rho_0$) from those of new
physics. Including the CDF top mass range such a separation became feasible
and from table~\ref{NPfits} we see that $\rho_0$ is remarkably close to unity,
leaving little room for any new physics which contributes to it. The allowed
regions in $\rho_0$ vs. $\hat{s}^2_Z$ from various observables and the
global fit are shown in Figure~\ref{rho}.
Similarly, $S_{\rm new}$, $T_{\rm new}$ and $U_{\rm new}$ are well-constrained
and consistent with
zero\footnote{Note, that the oblique parameters
are defined with a factor $\alpha$ factored out so that they are expected to
be of order unity if non-zero.}. The global fit yields a negative central value
for
$S_{\rm new}$, but consistent with 0 at 1 $\sigma$. This is in contrast
to $S < 0$, as was suggested by earlier data. ($A_{LR}^0$ by itself does
favor $S < 0$.) The allowed region in $S_{\rm new}$ and $T_{\rm new}$
is shown in Figure~\ref{ST}.

So far, we have allowed for new physics in the $Zb\bar{b}$ vertex only by an
overall factor in~(\ref{zbb}).  This implicitly assumes that the relative
contributions to the vector and axial vertices are such that there is little
effect on $A_{FB}^{0b}$. Indeed, $A_{FB}^{0b}$ can be seen in
Table~\ref{tblexp} to be in good agreement with the SM expectation.
However, one can do a more detailed analysis~\cite{SZ} by allowing separate
corrections to the left- and right handed couplings, i.e., the (lowest
order) couplings are replaced by
\be
\ba{l}
   g_{Lb} = {1\over 2} (g_{Vb} + g_{Ab}) \rightarrow
            - {1\over 2} + {1\over 3} \sin^2 \theta_W + \delta^b_L, \\
   g_{Rb} = {1\over 2} (g_{Vb} - g_{Ab}) \rightarrow
             {1\over 3} \sin^2 \theta_W + \delta^b_R.
\ea
\ee
{}From a global fit we obtain
\be
\ba{l}
   \delta^b_L = 0.0003 \pm 0.0047, \\
   \delta^b_R = 0.026 \pm 0.018
\ea
\ee
(with a correlation of 0.86). That is, $A_{FB}^{0b}$, which is consistent
with but slightly lower than the Standard Model prediction, forces the
$R_b$ anomaly to be in the right-chiral coupling.

\section{SUMMARY:}

The indirect determination of $m_t = 175 \pm 11^{+17}_{-19}$~GeV
(for $M_H = 300^{+700}_{-240}$~GeV) is in
spectacular agreement with the CDF range, $m_t = 174 \pm 16$~GeV,
while the somewhat lower value $160^{+11\; +6}_{-12\; -5}$ expected
in supersymmetry is still in reasonable agreement.
Also most other observables are in excellent agreement with the predictions of
the Standard Model. One exception is that the left-right asymmetry
is in direct conflict with LEP asymmetries. $\alpha_s$ determinations from
LEP (0.127(5)(2) from the lineshape, 0.123(6) from jets, and 0.122(5)
from $R_\tau$) are significantly higher than the ones performed at lower
energies.
The $Z \rightarrow b\bar{b}$ partial width exceeds the Standard Model
value by about 2.3 standard deviations. Interestingly, new physics which can
account for $\delta_{b\bar{b}}^{\rm new}$ would simultaneously decrease the
extracted $\alpha_s$ from $R$, bringing it in closer agreement with other
measurements. It remains to be seen to what extend these deviations
and the one in $A_{FB}^{0\tau}$ persist in the future.

Inclusion of the CDF result does not alter the Standard Model fits
significantly. It is, however, very useful in constraining new types
of physics: one can now separate the effects of new physics from $m_t$.
E.g, $\rho_0$, which describes sources of vector $SU(2)$
breaking beyond the SM, is now known to be very close to and consistent with
the SM value (of unity), $\rho_0 = 1.0012 \pm 0.0017 \pm 0.0017$.
The same is true for all the oblique parameters.
High precision experiments continue to prefer non-positive values for
$S_{\rm new}$ and a vanishing $T_{\rm new}$, but there is no longer a
significant indication of $S_{\rm new} < 0$. This is in contrast to
standard ETC/compositeness models, but is consistent with most of the parameter
space of minimal supersymmetry. The new data shows a slight preference
for a light Higgs mass close to the direct lower bound, but this is weak
statistically. One finds $M_H < 510$ (730) GeV at 90\% (95\%) c.l. It should be
kept in mind, however, that this limit depends almost entirely on
$R_b$ and $A_{LR}^0$, both of which are high compared to the SM.

\section*{Acknowledgements}
We are happy to thank Wolfgang Hollik, Bernd Kniehl and
Alberto Sirlin for discussions. This work was supported
by the Texas National Laboratory Research Commission and by the
D.O.E. under contract DE-AC02-76-ERO-3071. One of us (J.\ E.)
is supported in part by the Deutsche Forschungsgemeinschaft.

\vspace{4.0ex}


\onecolumn

\begin{table}
\centering
\caption{$Z$-pole observables from LEP and SLD and other recent measurements
compared to their Standard Model expectations. The Standard Model prediction
is based on $M_Z$ and uses the global best fit values for $m_t$
and $\alpha_s$, with 60 GeV $ < M_H < $ 1000 GeV. The fits include the
($M_Z$, $\Gamma_Z$, $R$, $\sigma^0_{\rm had}$, $A_{FB}^{0l}$) and
($R_b$, $R_c$; $\rho = -0.4$) correlations.}
\begin{tabular}{ccc}
Quantity & Value & Standard Model \\ \hline
$M_Z$ [GeV] & $91.1888 \pm 0.0044$ &  input \\
$\Gamma_Z$ [GeV] & $2.4974 \pm 0.0038$ & $2.497 \pm 0.001 \pm
0.003 \pm [0.002] $ \\
$R = \Gamma({\rm had})/\Gamma(l^+ l^-)$ & $20.795 \pm
0.040$ & $20.784 \pm 0.006 \pm 0.003 \pm [0.03]$ \\
$\sigma_{\rm had}^0 = \frac{12 \pi}{M_Z^2} \; \frac{\Gamma(e^+
e^-) \Gamma({\rm had})}{\Gamma_Z^2}$ [nb] & $41.49 \pm 0.12$
& $41.44 \pm 0.004 \pm 0.01 \pm [0.02]$ \\
$R_b = \Gamma(b \bar{b})/ \Gamma({\rm had})$ &$0.2202 \pm 0.0020$
& $0.2156 \pm 0 \pm 0.0004$ \\
$R_c = \Gamma(c\bar{c}) /\Gamma({\rm had})$ & $0.1583 \pm 0.0098$
& $0.171 \pm 0 \pm 0$ \\
$A_{FB}^{0l} = \frac{3}{4} \left( A_l^0 \right)^2$ &
$0.0170 \pm 0.0016$ & $0.0151 \pm 0.0005 \pm 0.0006$  \\
$A_{\tau}^0 \left(P_\tau \right)$ & $0.143 \pm 0.010$ & $0.142
\pm 0.003 \pm 0.003$ \\
$A_e^0 \left( P_\tau\right)$ & $0.135 \pm 0.011$ & $0.142 \pm
0.003 \pm 0.003$ \\
$A_{FB}^{0b} = \frac{3}{4} A^0_e A^0_b$ & $0.0967 \pm 0.0038$ &
$0.0994 \pm 0.002 \pm 0.002$ \\
$A_{FB}^{0c} = \frac{3}{4} A^0_e A^0_c$ & $0.0760 \pm 0.0091$ &
$0.071 \pm 0.001 \pm 0.001$ \\
$\bar{s}_e^2$ \ \ ($Q_{FB}$) & $0.2320 \pm
0.0016$ & $0.2322 \pm 0.0003 \pm 0.0004$ \\
$A_e^0 = A^0_{LR}$ \ \ (SLD) & $0.1637 \pm 0.0075
\;\; (92 + 93)$ & $0.142 \pm 0.003 \pm 0.003$ \\
     & $(0.1656 \pm 0.0076 \;\; (93))$ & \ \\
$N_\nu$ & $2.988 \pm 0.023$ & $3$ \\ \hline \hline \\ \\ \\ \\ \hline \hline
Quantity & Value & Standard Model \\ \hline
$M_W$ [GeV] & $80.17 \pm 0.18$ & $80.31 \pm 0.02 \pm 0.07$ \\
$M_W/M_Z$ (UA2) & $0.8813 \pm 0.0041$ & $0.8807 \pm 0.0002 \pm
0.0007$ \\
$Q_W$ (Cs) & $-71.04 \pm 1.58 \pm [0.88]$ & $-72.93 \pm 0.07 \pm
0.04$ \\
$g_A^{\nu e}$ (CHARM II) & $-0.503 \pm 0.017$ & $-0.506 \pm 0 \pm
0.001$ \\
$g_V^{\nu e}$ (CHARM II) & $-0.035 \pm 0.017$ & $-0.037 \pm 0.001
\pm 0$ \\
$s^2_W = 1 - \frac{M_W^2}{M_Z^2}$ & $0.2218 \pm 0.0059$ (CCFR) &
$0.2245 \pm 0.0003 \pm 0.0013$ \\
  & $0.2260 \pm 0.0048$ (All) & \\
$M_H$ [GeV] & $> 61$ (LEP) & $< \left\{ \begin{array}{l}
O (600) \; {\rm (theory)} \\ O (800) \; {\rm (indirect)} \end{array}
\right.$ \\
$m_t$ [GeV] & $> 131$ (D0) & $175 \pm 11^{+17}_{-19}$ (indirect) \\
  & $174 \pm 16$ (CDF) &  \\
$\alpha_s (M_Z)$ & $0.123 \pm 0.006$ (event shapes) &
  $0.127 \pm 0.005 \pm 0.002 \pm [0.001]$ \\
  & $0.116 \pm 0.005$ (event shapes + low energy) & ($Z$ lineshape) \\
\end{tabular}
\label{tblexp}
\end{table}

\begin{table}
\centering
\caption{Results for the electroweak parameters in the Standard
Model from various sets of data. The central values assume $M_H
= 300$~GeV, while the second errors are for $M_H \rightarrow 1000$ GeV (+) and
60 GeV ($-$). The last column is the increase in the overall $\chi^2$
to the fit as the Higgs mass increases from 60 to 1000 GeV.
The last two rows are the results of fits performed by the LEP
electroweak working group (LEP-EWG), with the appropriate translation
of $\bar{s}^2_e$ into $\hat{s}^2_Z$.}
\begin{tabular}{ccccc}
Set & $\hat{s}^2_Z$ & $\alpha_s (M_Z)$ & $m_t$ [GeV] & $\Delta
\chi^2_H$ \\ \hline
All indirect & $0.2317(3)(^1_2)$ & 0.127(5)(2) &
$175 \pm 11^{+17}_{-19}$ & 4.4 \\
Indirect + CDF ($174 \pm 16$) & $0.2317(3) (^2_3)$ & 0.127(5)(2) &
$175 \pm 9^{+12}_{-13}$ & 4.4 \\
Indirect + $\alpha_s$ ($0.116 \pm 0.005$) & $0.2316(3)(^1_2)$ & 0.122(3)(1) &
$178^{+10 \; +17}_{-11 \; -19}$ & 6.0 \\
LEP + low energy & 0.2320(3)(2) & 0.128(5)(2) &
$168^{+11 \; +17}_{-12 \; -19}$ & 2.7 \\
All indirect ($S=2.2$) & $0.2319 (3) (^1_2)$ & 0.128(5)(2) &
$170^{+11 \; +17}_{-12 \; -19}$ & 3.3 \\
$Z$-pole & 0.2316(3)(1) & 0.126(5)(2) &
$179^{+11 \; +17}_{-12 \; -19}$ & 4.2 \\
LEP & $0.2320 (4) (^1_2)$ & 0.128(5)(2) &
$170^{+12 \; +18}_{-13 \; -20}$ & 2.6 \\
SLD + $M_Z$ & 0.2291(10)(0) & --- &
$251^{+24 \; +21}_{-26 \; -23}$ & --- \\ \hline
$Z$-pole, $M_W$, recent $\nu$ (LEP-EWG) & $0.2317(3) (^0_2)$ & 0.125(5)(2) &
$178 \pm 11^{+18}_{-19}$ & \\
LEP (LEP-EWG) & $0.2319(4) (^1_2)$ & 0.126(5)(2) &
$173^{+12 \; +18}_{-13 \; -20}$ & \\
\end{tabular}
\label{tblres}
\end{table}

\begin{table}
\squeezetable
\centering
\caption{Results for the electroweak parameters including additional fit
parameters describing physics beyond the SM. All fits include the CDF
constraint $m_t = 174 \pm 16$ GeV. The central values
are for $M_H = 300$ GeV, the upper second errors for $M_H = 1000$ GeV
and the lower ones for $M_H = 60$ GeV. For $T_{\rm new}$ we also list the
equivalent $\rho_0 \equiv 1 + \alpha T_{\rm new}$.}
\begin{tabular}{ccccccc}

$\hat{s}^2_Z$ & $\alpha_s (M_Z)$ & $m_t$ [GeV] & $S_{\rm new}$ &
$T_{\rm new}$ & $U_{\rm new}$ & $\delta_{b\bar{b}}^{\rm new}$ \\
 & & & & ($\rho_0$) & & \\
\hline

0.2317(3)(3) & 0.127(5)(2) & $175 \pm 9^{+12}_{-13}$ &
--- & --- & --- & --- \\

0.2316(3)(2) & 0.111(9)(0) & $177 \pm 9 \pm 13$ &
--- & --- & --- & $0.023 \pm 0.011 \pm 0.003$ \\

0.2316(3)(1) & 0.125(6)(1) & $166 \pm 15 \pm 0$ &
--- & $0.16 \pm 0.23 \pm 0.23$ & --- & --- \\
    & & & & (1.0012(17)(17)) & & \\

0.2316(3)(2) & 0.111(9)(0) & $174 \pm 16^{+1}_{-0}$ &
--- & $0.05 \pm 0.25 \pm 0.25$ & --- & $0.022 \pm 0.011 \pm 0$ \\
    & & & & (1.0004(18)(18)) & & \\

0.2314(4)(1) & 0.125(6)(0) & $167 \pm 15 \pm 0$ &
$-0.21 \pm 0.24^{-0.08}_{+0.17}$ & $0.03 \pm 0.30^{+0.17}_{-0.10}$ & $-0.50 \pm
0.61$ &
--- \\
    & & & & ($1.0002(22)(^{12}_{\; 7})$) & & \\

0.2313(4)(1) & 0.112(9)(0) & $175 \pm 16 \pm 0$ &
$-0.21 \pm 0.24^{-0.08}_{+0.17}$ & $-0.09 \pm 0.32^{+0.16}_{-0.11}$ & $-0.53
\pm 0.61$ &
$0.022 \pm 0.011 \pm 0$ \\
    & & & & ($0.9993(23)(^{12}_{\; 8})$) & & \\

\end{tabular}
\label{NPfits}
\end{table}


\clearpage
\pagestyle{empty}

\begin{figure}[h]
\vspace{40mm}
\postscript{fig1.ps}{0.8}
\caption{Increase in $\chi^2$ from the best fit for all data, with and
without the CDF constraint $m_t = 174 \pm 16$ GeV, and distributions
omitting $R_b$ and/or $A_{LR}$.}
\label{Higgs}
\end{figure}

\begin{figure}[h]
\vspace{40mm}
\postscript{fig2.ps}{0.8}
\caption{Allowed regions at 90\% c.l. in $\rho_0$ and $\hat{s}^2_Z$
from the $Z$-pole asymmetries; the $Z$ widths (and CDF $m_t$); and
$M_Z$, $M_W$ and $m_t$, assuming $M_H = 300$ GeV. Also shown are the
allowed regions for all data, including the CDF $m_t$, for $M_H
= 60$, 300, and 1000 GeV. }
\label{rho}
\end{figure}

\begin{figure}[h]
\vspace{40mm}
\postscript{fig3.ps}{0.8}
\caption{Allowed regions at 90\% c.l. in $S_{\rm new}$ and
$T_{\rm new}$ for various observables. The fit to $M_Z$, $M_W$
assumes $U_{\rm new} = 0$, but $U_{\rm new}$ is left as free parameter
in the other cases. The fits to all data are shown for $M_H = 60$,
300, and 1000 GeV. The CDF $m_t$ constraint is always included in
the fits.}
\label{ST}
\end{figure}

\end{document}